\documentclass[aps,prl,twocolumn,superscriptaddress]{revtex4}

\usepackage[dvips]{graphicx}
\usepackage{amsmath}
\usepackage{subfigure}

\renewcommand{\Vec}[1]{{\bf #1}}

\begin{document}


\title{Optical Hall conductivity in ordinary and graphene QHE systems}



\author{Takahiro Morimoto}
\affiliation{Department of Physics, University of Tokyo, Hongo, 
Tokyo 113-0033, Japan}
\author{Yasuhiro Hatsugai}
\affiliation{Institute of Physics, University of Tsukuba, Tsukuba, 
305-8571, Japan}
\author{Hideo Aoki}
\affiliation{Department of Physics, University of Tokyo, Hongo, 
Tokyo 113-0033, Japan}


\date{\today}

\begin{abstract}
We have revealed from a numerical study that 
the optical Hall conductivity $\sigma_{xy}(\omega)$ has 
a characteristic feature even in the ac ($\sim$ THz) regime 
in that the Hall plateaus are retained 
both in the ordinary two-dimensional electron gas and in graphene 
in the quantum Hall (QHE) regime, 
although the plateau height is no longer quantized in ac.  
In graphene $\sigma_{xy}(\omega)$
reflects the unusual Landau level structure.  
The effect remains unexpectedly robust 
against a significant strength of disorder, which we attribute 
to an effect of localization.  
We predict the ac quantum Hall measurements are feasible through the 
Faraday rotation characterized by the fine-structure constant $\alpha$.
\end{abstract}


\maketitle


{\it Introduction ---} 
There is a continuing fascination with the 
quantum Hall effect (QHE), despite its long history, 
along various avenues.  
While most of the works have concentrated on static properties, 
one direction that has not been fully explored is 
optical properties of the quantum Hall system, 
which is exactly the purpose of the present study.  
One motivation comes from the fact that recent experimental advances 
in spectroscopy 
in the THz regime are making optical measurements a reality for QHE systems 
with the relevant energy scale being THz in magnetic fields 
of a few tesla.\cite{hangyo,ikebe2008cds}  
To be more precise, these authors have 
observed ellipticity and Faraday rotation in a usual QHE in 
a two-dimensional electron gas (2DEG), 
and found a resonance structure at a cyclotron frequency $\omega_c 
\sim $ THz from the Hall angle, 
$
\Theta_H(\omega)=\frac{1}{2}\arg (t_+(\omega)/t_-(\omega)),
$
which is directly connected to optical conductivity, since 
the transmission coefficients, 
$
t_{\pm}(\omega)=2n_0/[n_0+n_s+\sigma_{\pm}(\omega)/(c\varepsilon_0)],
$
is related to the optical conductivity for circularly polarized light via 
$\sigma_{\pm}(\omega)=\sigma_{xx}(\omega)\pm i \sigma_{xy}(\omega)$\cite{oconnell82,gusynin-sumrule}.

Another motivation of the present study is 
the recent emergence of the physics of graphene, where the 
anomalous QHE specific to the ``massless Dirac" 
electrons\cite{Nov05,Kim-gr} is attracting keen interests.  
So the second purpose of the Letter is to study $\sigma_{xy}(\omega)$ for graphene in the QHE regime as compared with those 
in the ordinary 2DEG.  For graphene, optical properties begin to be studied: 
the longitudinal optical conductivity $\sigma_{xx}(\omega)$ 
has been measured through the transmission \cite{sadowski}, or 
theoretically examined in terms of the cyclotron emission\cite{morimoto-CE}. 
Here we look for features in the optical {\it Hall} conductivity $\sigma_{xy}(\omega)$ in graphene.

Theoretically, the question is how the static quantum Hall effect, 
a topological phenomenon\cite{tknn,hatsugai1993cna}, should evolve into the optical Hall conductivity 
$\sigma_{xy}(\omega)$ in the ac regime.  
Naively, one might expect the plateau 
structure in the Hall conductivity may be immediately washed out 
as we go into an ac regime where the topological protection no 
longer exists.  To explore whether this intuition holds, 
here we have calculated the 
optical Hall conductivity in ordinary and graphene QHE systems 
to probe the ac quantum Hall physics, where the conductivity is 
calculated from Kubo formula with 
a numerical (exact diagonalization) method, since we want to 
incorporate effects of Anderson localization.  
We start with the ordinary 2DEG, since even for 2DEG the ac 
conductivity has only been dealed with a phenomenological 
(Drude) formalism\cite{oconnell82} or with Maxwell's equations\cite{peng1991frq}. 

We shall show from the numerical study that (i) the plateau structure in the 
QHE in 2DEG is retained, up to significant degree of disorder, 
even in the ac (THz) regime, although the heights of the plateaus 
are no longer quantized in the ac regime.  
We attribute the unexpected robustness to an effect of localization, 
where the existence of extended states and mobility 
gaps between them ensure 
the step structures in the ac Hall conductivity.
(ii) For graphene, the optical Hall conductivity 
reflects the unusual Landau level structure.  
(iii) We then 
predict the ac quantum Hall effect can be detected through Faraday 
rotation measurements as a step structure in $\Theta_H(\omega)$,
whose magnitude is estimated to be of the order of the fine-structure constant $\alpha (\sim7$ mrad), which is within the experimental feasibility.  
If one utilizes a free-standing graphene, for which 
$\alpha$ has been seen as transparency\cite{nair-alpha}, 
the rotation angle should be
exactly $\alpha$.

A word about the nature of the random potential here: 
it has long been known that the effect of the Anderson localization 
on the (static) conductivity is qualitatively different between 
short-range and long-range scatters in 2DEG\cite{huckestein}.  
Long-range potential should also be relevant to the ripples in graphene.
One particular interest, in the static graphene QHE, is that graphene has an anomalously robust 
$n=0$ Landau level and the associated 
QHE step against disorder when it is slowly varying,
which is related to 
topologically protected Atiyah-Singer's theorem.\cite{Nov05,nomura2008qhe,guinea-ripple,kawarabayashi09}
So we here examine $\sigma_{xy}(\varepsilon_F,\omega)$ 
for long-range scatterers in the exact diagonalization study 
that takes care of the localization effects, 
where we have a question  in mind --- 
what will become of the Hall plateau structure in the ac regime, 
especially for the $n=0$ Landau level in graphene.

{\it Optical Hall conductivity in ordinary QHE ---} 
Let us first look at the optical Hall conductivity 
in the QHE in 2DEG 
as realized in GaAs/AlGaAs with a Hamiltonian,
$H_0=\frac{1}{2 m^*}(\Vec{p} +e \Vec{A})^2 , $
and the current matrix elements, 
$
j_{x}^{n,n'}=
i\sqrt{\hbar \omega_c/2m^*}\left(\sqrt{n}\delta_{n-1,n'}-\sqrt{n+1}\delta_{n+1,n'}\right),$
$
j_{y} ^{n,n'}=\sqrt{\hbar \omega_c/2m^*}\left(\sqrt{n}\delta_{n-1,n'}+\sqrt{n+1}\delta_{n+1,n'}\right),
$
where $n, n'$ are the Landau indices and $\omega_c$ the cyclotron frequency.  
 
To include the effect of disorder, we employ the exact diagonalization method for the disorder 
described by randomly placed scatterers with a potential 
$V(\Vec r)=\sum_j u_j \exp(-|\Vec r-\Vec R_j|^2/2d^2)/(2\pi d^2),$
where $d$ is the range of the potential, while the strength of the potential, 
each placed at $\Vec R_j$, is assumed to take $u_j=\pm u$ with  
random signs so that the density of states broadens symmetrically in energy.  
We adopt $d=0.7\ell$, which is comparable to  the magnetic length $\ell=\sqrt{\hbar/eB}$.  
The degree of disorder can be characterized by 
$\Gamma^2/4=u^2 N_{\rm imp}/[2\pi(\ell^2+d^2)L^2]$ 
with $N_{\rm imp}$ being the number of impurities, where $\Gamma$ 
measures the Landau level broadening.\cite{ando,morimoto-CE} 
\footnote{
It is known that the presence of the plateau structure in the integer QHE 
does not depend on the details of disorder, 
although the nature of disorder,  such as long-raged/short-ranged 
or chiral symmetric or not, can affect the detail of the structure.}
Diagonalization of the Hamiltonian is done by
retaining 7 Landau levels.
We have numerically checked that this is sufficient
in the energy range considered here 
for the linear sample dimension of $L= 15\ell$.
For the ensemble average
we have taken 5000 random configurations.

To calculate the optical conductivity in the QHE system we 
use the Kubo formula, 
\begin{equation}
\sigma_{xy}(\omega)
=
\frac{i\hbar e^2}{ L^2}
\hspace{-0.9em}
\sum_{
\begin{array}{c}
\scriptstyle \epsilon_a < \varepsilon_F \\[-0.4em] 
\scriptstyle \epsilon_b \ge \varepsilon_F\\[-1.3em] 
\end{array}
}
\hspace{-0.9em}
\frac 1 {\epsilon_b-\epsilon_a}
\left(\frac{j_x^{ab}j_y^{ba}}{\epsilon_b-\epsilon_a-\hbar\omega} 
-\frac{j_y^{ab}j_x^{ba}}{\epsilon_b-\epsilon_a+\hbar\omega}
\right) ,
\label{kuboformula}
\end{equation}
where $\epsilon_a$ is the eigenenergy, $j_x^{ab}$ the current matrix 
elements between the eigenstates, and $\varepsilon_F$ the Fermi energy.

\begin{figure}[tbp]

\begin{tabular}{c}
\begin{minipage}{\linewidth}
\includegraphics[width=\linewidth]{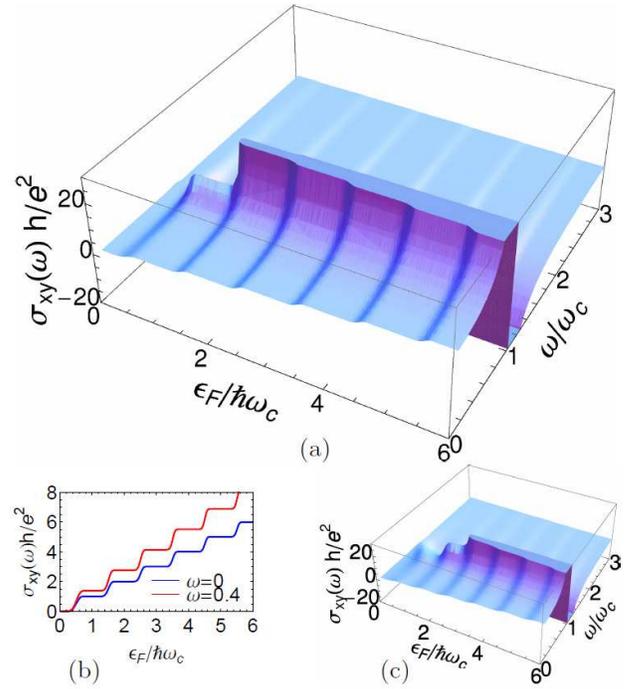}
\end{minipage}
\end{tabular}
\caption{
(Color online) 
Exact diagonalization result for 
(a) the optical Hall conductivity $\sigma_{xy}(\varepsilon_F,\omega)$ with $\Gamma=0.2 \hbar\omega_c$, 
(b) static (blue) and optical (red) Hall conductivity $\sigma_{xy}(\varepsilon_F,\omega)$, and
(c) $\sigma_{xy}(\varepsilon_F,\omega)$ with larger disorders $\Gamma=0.5 \hbar\omega_c$
in usual QHE system.
}
\label{longrange-gaas}
\end{figure}

Figure \ref{longrange-gaas} shows the results for the usual QHE system.
We plot 
$\sigma_{xy}$ on an $(\omega, \varepsilon_F)$ plane, 
where the $\omega=0$ cross section corresponds to the 
familiar static QHE.  
We immediately notice two features: 
(i) $\sigma_{xy}(\omega)$ for a fixed value of $\varepsilon_F$ 
exhibits a resonance structure around 
the cyclotron frequency (as observed in the experiment\cite{hangyo}).  
(ii) Away from the resonance, {\it a step-like structure} is preserved in $\sigma_{xy}(\omega)$ as a function of $\varepsilon_F$ 
for each value of $\omega$.  
Although the step heights are not quantized exactly,
the flatness is surprisingly preserved as seen in 
Fig.\ref{longrange-gaas}(b).    
If we first look at the clean limit, we can rewrite eqn(\ref{kuboformula}) as 
$\sigma_{xy}(\omega)
\rightarrow n \frac{e^2}{h}\frac{\omega_c^2}{\omega_c^2-\omega^2},$ 
since we can replace $\varepsilon_{n+1}-\varepsilon_n$ 
with $\hbar\omega_c$ for 
$\varepsilon_F$ between $n$ and $n+1$ Landau levels, 
which shows a resonance structure around $\omega \simeq \omega_c$.  

The step structure is in fact a quantum effect (outside 
the Drude picture).  In the dc QHE, the localization is the cause of the 
plateaus in the Aoki-Ando picture\cite{aoki-ando1981}. 
In the ac QHE, the Kubo formula, eqn(1), contains $\omega$, 
and does not simply reduce to a topological expression.
In this sense the result for the robust plateaus is quite nontrivial. 

The physical insight for the unexpectedly robust ac Hall step structure
is that the main contribution to the optical Hall conductivity comes
from the delocalized states 
whose existence ensures the robust step structure in ac Hall conductivity.
To be more precise, the magnitude of the current matrix elements in eqn(\ref{kuboformula}) is much larger for the extended states than for localized 
states, so that the optical Hall conductivity is dominated by the transitions 
between the extended states 
which reside around the center of each Landau level, while 
the localized states give rise to the step structure.  
Thus the message here is that the existence of 
localized  and extended states manifests itself as 
step structures even in the ac regime.

{\it Optical Hall conductivity in graphene ---} 
We now turn to the optical Hall conductivity in graphene.
Here we again adopt the exact diagonalization method for 
the disorder potential introduced by randomly placed scatterers.  
When the range of the random potential 
is much larger than the lattice constant in graphene, the 
scattering between K and K' points in the Brillouin zone is suppressed, 
so that we can assume the random term takes a 
diagonal form in the Dirac Hamiltonian as
\begin{equation}
H_0+V=v_F 
\begin{pmatrix}
0 & \pi^- &0&0 \\
\pi^+ &0&0&0 \\
 0&0&0 & \pi^+ \\
 0&0& \pi^- &0 \\ 
\end{pmatrix}
+
\begin{pmatrix}
1&0&0&0\\
0&1&0&0\\
0&0&1&0\\
0&0&0&1\\ 
\end{pmatrix}
V(\Vec r)
,
\label{longrange-ham}
\end{equation}
So we adopt the Dirac model as in Refs.\cite{nomura-ryu-koshino,nomura2008qhe} 
to obtain wave functions and conductivity in the presence of disorder.
Retaining 9 Landau levels with system size $L=15\ell$, 
we have calculated $\sigma_{xy}(\varepsilon_F,\omega)$ with the Kubo formula eqn(\ref{kuboformula}) with the current matrix 
for graphene\cite{ZhengAndo,morimoto-CE}.

\begin{figure}[tbp]

\begin{tabular}{c}
\begin{minipage}{\linewidth}
\includegraphics[width=\linewidth]{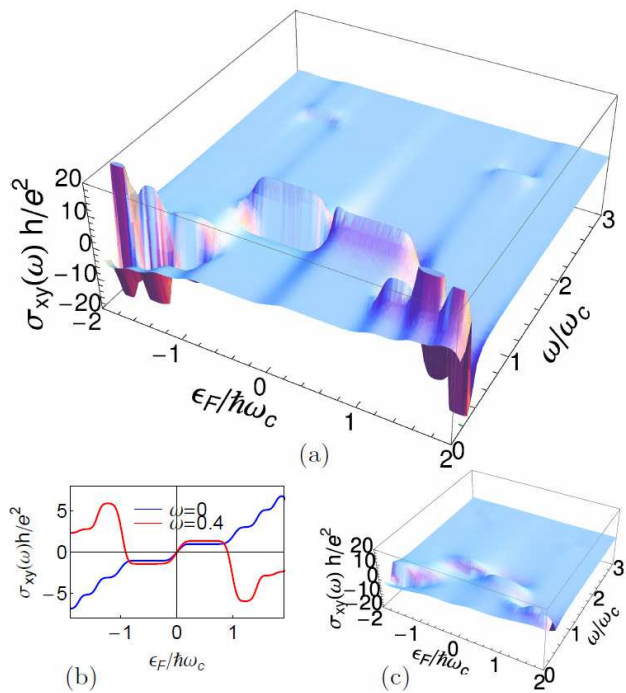}
\end{minipage}
\end{tabular}
\caption{
(Color online) 
Exact diagonalization result for 
(a) the optical Hall conductivity $\sigma_{xy}(\varepsilon_F,\omega)$ with $\Gamma=0.2 \hbar\omega_c$, 
(b) static (blue) and optical (red) Hall conductivity $\sigma_{xy}(\varepsilon_F,\omega)$, and
(c) $\sigma_{xy}(\varepsilon_F,\omega)$ with larger disorders $\Gamma=0.5 \hbar\omega_c$
for the graphene QHE system.
}
\label{longrange}
\end{figure}

In the result, Fig.\ref{longrange}, 
we notice several features distinct from the 
result for the ordinary QHE system (Fig.\ref{longrange-gaas}):  

(i) The optical Hall conductivity 
$\sigma_{xy}(\varepsilon_F,\omega)$ exhibits a more complex 
structure, which reflects the Landau levels, 
${\rm sgn}(n)\sqrt{|n|}\hbar \omega_c$ with $\omega_c=v_F\sqrt{2eB/\hbar}$, 
that are not uniformly spaced for the massless Dirac dispersion.  
Thus a series of resonances appear around 
many allowed transitions, $|n|-|n'|=\pm1$.  More precisely, 
the low-$\omega$ resonances appear as $n \to n+1$ transitions, 
while resonances ($-n \to n+1$) across the Fermi point emerge 
in larger $\omega$ region. 

(ii) Away from these resonances, we again observe that 
{\it step-like structures} remain in the optical Hall conductivity, 
as clearly seen in Fig.\ref{longrange}(b).  
Due to the electron-hole symmetry, $\sigma_{xy}(\varepsilon_F, \omega)$ is 
odd in $\varepsilon_F$ throughout, so the step structure is symmetric as well.

If we more closely look at the result, 
while the ac Hall steps for larger values of $|n|$ are smeared for smaller values of 
$\omega$ because different Landau levels sit close to each other, 
the ac Hall steps for small values of $n (=0, \pm 1)$ 
are robust.  
Specifically, the $n=0$ step remains up to $\Gamma$ as large 
as $0.7\hbar\omega_c$.  
One reason should be the $n=0$ Landau level stands alone, 
but another one we note is the electron-hole symmetry.  Namely, 
the self-energy (arising from the randomness) 
in Green's function contains off-diagonal elements 
between $\pm n$ Landau levels, where 
$n \neq 0$ states are significantly 
affected by this effect, while the $n=0$ Landau level has itself as the 
electron-hole partner with no off-diagonal element.  
In terms of Hamiltonian, graphene QHE is square root of the usual QHE, 
so that $n$-th Landau level in usual QHE bifurcates into $\pm \sqrt n$ Landau levels in graphene but not for $n=0$.

We note in passing that the present result in the exact diagonalization 
differs from what we would have with the 
self-consistent Born approximation(SCBA)\cite{morimoto-LT-opthall}, 
where the Landau level broadening 
and the plateau-to-plateau transition width for different 
Landau indices are similar $\sim \Gamma$
\cite{ZhengAndo}, 
since localization is not considered.

\begin{figure}[tbp]
\begin{center}
\begin{minipage}{\linewidth}
\subfigure[]{\includegraphics[width=\linewidth]{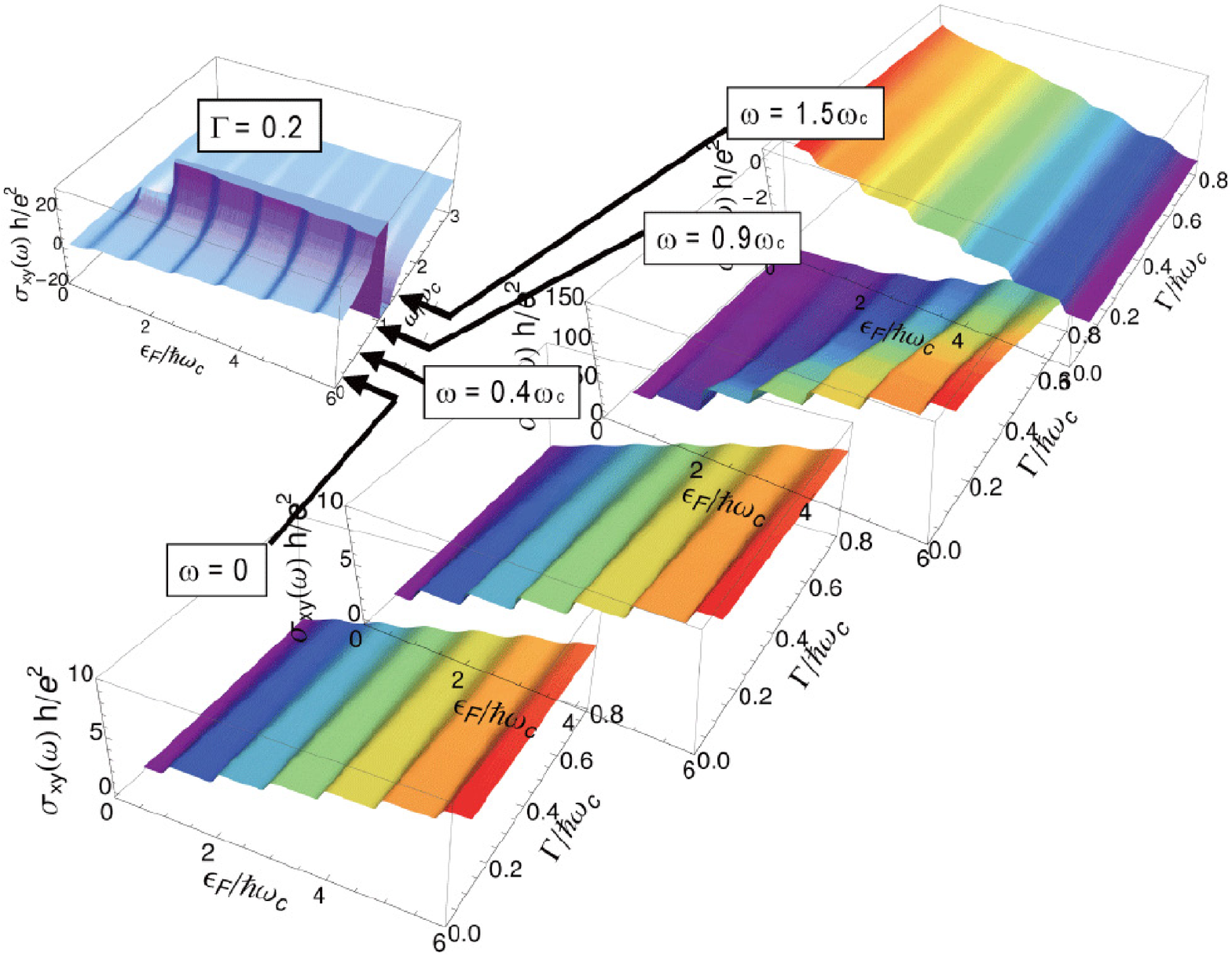}}
\subfigure[]{\includegraphics[width=\linewidth]{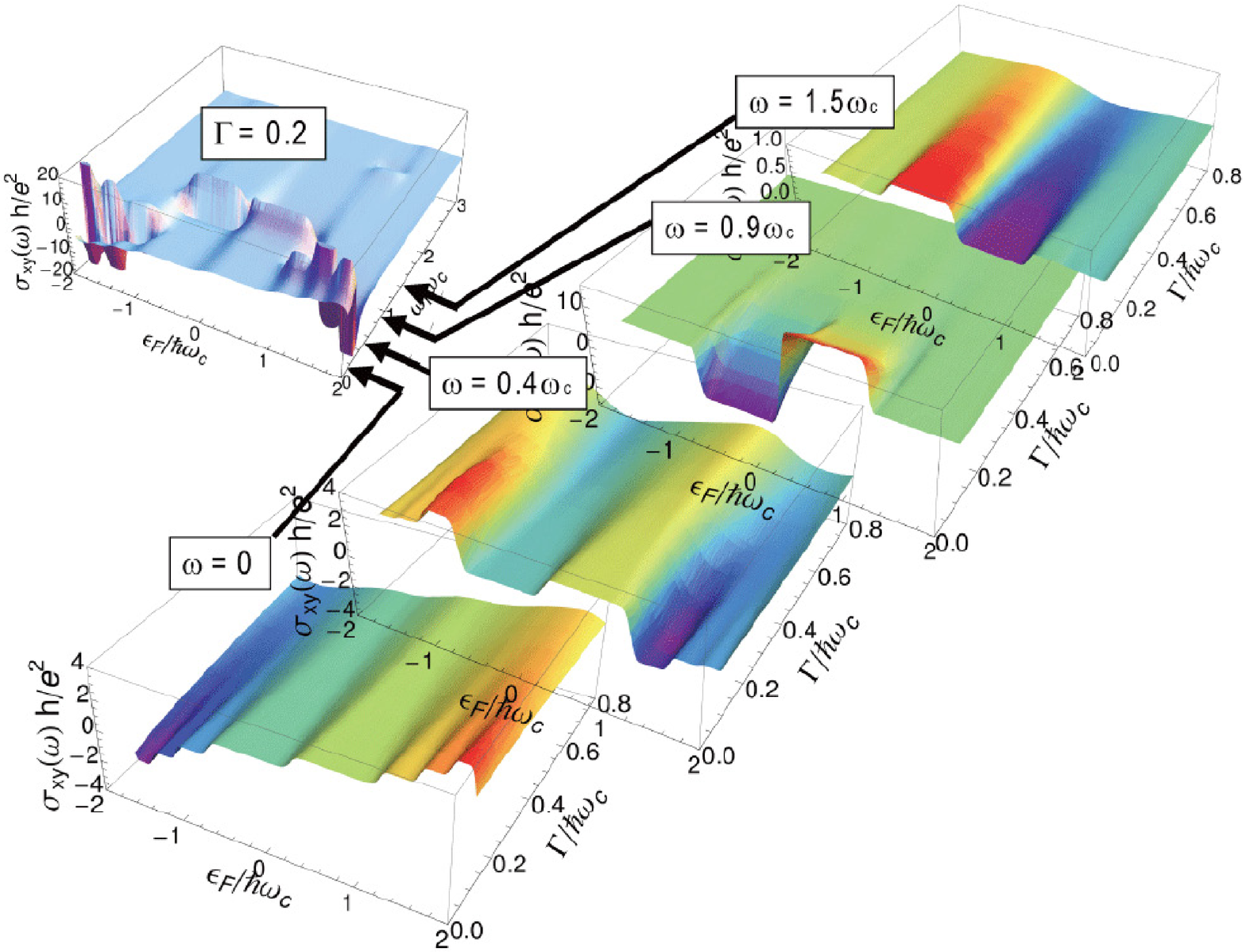}}
\end{minipage}
\end{center}
\caption{
(Color online) 
Exact diagonalization result for the optical Hall conductivity, 
$\sigma_{xy}(\varepsilon_F,\omega)$, plotted against Fermi energy and disorder strength $\Gamma$ for various values of frequency $\omega = 0, 0.4\omega_c, 
0.9\omega_c, 1.5\omega_c$ in (a)the ordinary and (b)graphene QHE system.
}
\label{gdep}
\end{figure}

{\it Robustness of the step structure ---} 
We finally examine how the step-like structure in the optical Hall 
conductivity vanishes as we further increase the degree of disorder. 
So we have calculated $\sigma_{xy}(\varepsilon_F,\omega)$ against 
the strength of disorder, $\Gamma$, for each value of $\omega$ 
with exact diagonalization.  
For the ordinary QHE system in 
Fig.\ref{gdep}, we can see that the step structure remains 
up to $\Gamma \simeq 0.7\hbar\omega_c$ for each value of $\omega$, 
both below and above the cyclotron resonance.  
While the density of states (not shown) broadens with a width $\sim \Gamma$, 
the step structure in the optical Hall conductivity $\sigma_{xy}(\omega)$ 
is blurred with $\Gamma$ 
much more slowly.  
We can also notice in the result for $\omega=0.9\omega_c$, very close to 
the cyclotron resonance where the ac Hall conductivity exceeds 
100 times $e^2/h$, that the step structure is surprisingly preserved in such a 
resonant region.

Let us move on to graphene QHE (Fig.\ref{gdep}).
The step structure for $n \neq 0$ in the optical Hall conductivity 
$\sigma_{xy}(\varepsilon_F,\omega)$ 
is less robust against disorder 
than in 2DEG, again due to the multiple cyclotron resonance 
frequencies.  However, the step corresponding to $n=0$ Landau level 
is robust unless $\omega$ is too close to a resonance.  
In the static Hall conductivity $\sigma_{xy}(\omega=0)$,
 $n\neq 0$ Hall steps are smeared as soon as Landau levels are merged 
while the step associated with $n=0$ Landau level 
is robust, 
which indicates that the extended states in $n=0$ Landau level are 
unusually robust.\cite{nomura2008qhe,kawarabayashi09}  
The present ac result indicates that 
the step in $\sigma_{xy}(\varepsilon_F,\omega)$ associated with $n=0$ Landau level exhibits special robustness against disorder in the ac regime as well, 
which we take to be the effect of localization and
the electron-hole symmetry.

The topological formulation of the static Hall conductivity relies on 
gaps between the mobility edges.  
The mobility gap structure 
is considered to bring about the robust step strucutre
in the ac region if
the frequency $\hbar\omega$ is smaller than 
$\hbar\omega_c$ (the energy spacing between the delocalized 
states).   
The present result (Fig.\ref{gdep})
does indicate that the robust step structure survives 
most prominently for $\omega<\omega_c$.  
In this sense the topological structure
associated with extended states remains in the ac regime.

{\it Faraday rotation ---} 
To summarize, we have revealed that the  
optical Hall conductivity in both the ordinary QHE and graphene QHE systems 
has plateau structures that persist even in ac regimes 
for significant strengths of disorder. 
Finally let us mention the experimental feasibility.  
We propose that the ac Hall steps should be 
observable through accurate Faraday-rotation measurements in the THz to 
far-infrared spectroscopy.  
This is because the Faraday-rotation angle $\Theta_H$ is directly connected to 
the optical Hall conductivity via
$
\Theta_H =
\frac{1}{2}\arg \left(\frac{t_+(\omega)}{t_-(\omega)}\right)
\sim
\frac{1}{(n_0+n_s)c\varepsilon_0}\sigma_{xy}(\omega),
$
where $n_0 (n_s)$ is the refractive index of air (substrate), 
and we have assumed $n_0+n_s\gg\sigma_{\pm}/(c\varepsilon_0)$ in 
the last line. 
Hence the Faraday rotation angle is proportional 
to $\sigma_{xy}(\omega)$, 
so that the step structure in $\sigma_{xy}(\omega)$ should be observed 
as jumps in Faraday-rotation measurements.  
We can estimate the size of the 
jumps $\Delta\Theta_H$ by putting $\sigma_{xy}\sim e^2/h$ (when 
$\omega$ is well below the resonance), so that
\begin{eqnarray}
\Delta\Theta_H
\sim
\frac{1}{(n_0+n_s)c\varepsilon_0}\frac{e^2}{h}
\sim
\frac{2}{n_0+n_s}\alpha
\quad\sim
7 \ \mbox{mrad},
\end{eqnarray}
where $\alpha=e^2/(4\pi\varepsilon_0\hbar c)$ is the fine-structure constant. 
So the steps in the Faraday-rotation angle should be 
of the order of the fine structure constant.  Recently Shimano {\it et al.} have achieved an experimental resolution of $\sim 1$ mrad 
\cite{ikebe2008cds}, so that the present 
effect is well within the experimental feasibility.  
While Nair et al.\cite{nair-alpha} have seen the fine-structure constant 
from visual transparency of graphene, the proposal here 
amounts to the fine-structure constant seen from a rotation.  

One future problem is 
how we can capture $\sigma_{xy}(\omega)$ in terms of the 
dynamical scaling argument.  
There are literatures on the dynamical scaling for 
$\sigma_{xx}(\omega)$ and associated dynamical critical 
exponent\cite{gammel-brenig},
so an extension to $\sigma_{xy}(\omega)$ should be interesting,
since we have noticed that the step structure becomes 
slightly sharper when we go from the sample size $L=10\ell$ to $15\ell$.

We wish to thank Ryo Shimano, Yohei Ikebe, Seigo Tarucha, Takeo Kato and Takashi Oka for 
illuminating discussions.  
This work has been supported in part by Grants-in-Aid for Scientific Research, 
No.20340098 (YH, HA), 20654034 (YH) from JSPS and 
Nos.220029004, 20046002 (YH) from MEXT.

\end{document}